\documentclass[aps,nofootinbib,prd,eqsecnum,showpacs,showkeys,preprintnumbers]{revtex4-2}
\usepackage[caption=false]{subfig}
\usepackage{graphicx}
\usepackage{amsmath}
\usepackage{amsfonts}
\usepackage{amssymb}
\usepackage{float}
\usepackage{color}
\usepackage{bm}
\usepackage{mathrsfs}
\usepackage{epstopdf}
\usepackage{url}
\usepackage{footnote}
\usepackage{textcomp}
\usepackage{ulem}
\usepackage{esint}
\usepackage[unicode=true, pdfusetitle,
 bookmarks=true,bookmarksnumbered=false,bookmarksopen=false,
 breaklinks=false,pdfborder={0 0 1},backref=false,colorlinks=false]{hyperref}
\usepackage{multirow}
\usepackage{pifont}

\begin{document}

\title{Observational constraints on Tachyon inflation and reheating in $f(Q)$
gravity}
\author{K. El Bourakadi$^{1,2}$}
\email{k.elbourakadi@yahoo.com}
\author{Z. Sakhi$^{1,2}$}
\email{zb.sakhi@gmail.com}
\author{M. Bennai$^{1,2}$}
\email{mdbennai@yahoo.fr}
\date{\today }

\begin{abstract}
In this work we study one of the most appealing string theory motivated\
models, we present a tachyonic inflationary model in the recently proposed
symmetric teleparallel framework, and examine constraints on tachyon
inflation with the exponential potential along with the reheating for a
chosen $f(Q)$ gravity model. Considering a reheating phase parametrized by a
number of e-folds $N_{re},$ a temperature $T_{re}$, and an equation of state 
$\omega _{re}$, we relate the reheating parameters as functions of the
exponantial tachyon potential, $f(Q)$ model, and\ the spectral index $n_{s}$
parameters. We argue that our model predicts inflationary e-folds bounded as 
$50\leq N\leq 64$. While for the reheating phase, wide ranges of reheating
e-folds numbers and temperatures can be obtained as we increase $\omega
_{re} $\ towards the value $1/4$ according to recent Planck Data.\ 
\end{abstract}

\affiliation{$^{1}${\small Quantum Physics and Magnetism Team, LPMC, Faculty of Science
Ben M'sik,}\\
{\small Casablanca Hassan II University, Morocco}\\} 
\affiliation{$^{2}$LPHE-MS Laboratory department
of Physics, Faculty of Science, Mohammed V University in Rabat, Morocco.\\}

\maketitle

\section{Introduction}


Inflation \cite{I1} presents the appealing prospect of resolving many of the
mysteries of standard hot big bang cosmology. A period of slow-roll
evolution of the inflaton scalar field requires that the potential energy $%
V(\phi )$ corresponding to inflation dominates over the kinetic term $\dot{%
\phi}^{2}/2$ and drives a quasi-exponential expansion of the Universe, which
is a critical condition to successful inflation. The inflationary scenario
generates seeds for cosmic microwave background (CMB) anisotropy \cite%
{I2,I3,I4}, which predicts Gaussian perturbations that could be\ validated
by observation \cite{I5,I6,I7,I8}. Moreover, primordial density
perturbations from the early Universe are caused by quantum fluctuations in
the scalar field that are coupled to metric fluctuations. During inflation,
there are also vacuum fluctuations of the metric that produce primordial
gravitational waves \cite{I9,I10,I11,IY,IYY,IY3,IY4,IY5,IY6}. The potential
corresponding to inflation can\ determine the power spectra of primordial
density perturbations and tensor perturbations created during that phase,
which may be calculated using particle physics models or string theory.
However, there is no favored concrete inflationary scenario based on a
credible realistic particle physics model at the moment. String theory, on
the other hand, offers many weakly coupled scalar fields that could be
candidates for\ specific models of inflation. However, their
non-perturbative potentials $V(\phi )$ do not appear suitable to maintain
slow-roll inflation because for large values of $\phi $, they either grow or
tend to zero too quickly. As a result, it is critical to investigate novel
options for incorporating an inflationary evolution of the early Universe.

K-inflation is an alternative to standard inflationary models \cite{I12,I13}%
, in which inflation is accomplished through a nonstandard kinetic component
in the inflaton's Lagrangian. Tachyon inflation \cite{I14} is an appealing
and popular model of K-inflation that can be implemented in Type-II string
theory. In open string theory, Ref. \cite{I15} has built a classical
time-dependent solution that characterizes the decay process of an unstable
D-brane. During the decay process, the tachyonic field on the brane rolls
down toward the potential minimum. There has been much research on the
different cosmological consequences of the rolling tachyon \cite{I16,I17,I18}%
. In general, the unusual nature of the tachyonic action distinguishes
cosmology with a tachyonic field from cosmology with a standard scalar field.

Many efforts to explain the observational evidence of cosmic acceleration
gives a new view on current cosmology. The phenomenon is known as Dark
Energy (DE), however, it is far from easy to reconstruct a coherent and
self-consistent cosmic history that works at every epoch. The cosmological
constant is a simple explanation that should work in both the early
inflation and late dark energy epochs. The approach leads us to the $\Lambda 
$CDM model, which claims that the universe is made up of $72\%$ DE, $24\%$
Dark Matter (DM), and $4\%$ visible matter, necessitating an early
inflationary epoch capable of addressing the shortcomings of the
Cosmological Standard Model based on General Relativity, Big Bang
Nucleosynthesis, and Standard Model of Particles. Despite its
accomplishments, this model suffers from critical confusion caused by a
large disparity between the value anticipated by any quantum gravity theory
and the current observable value. Moreover, no conclusive experimental
evidence of new particles capable of understanding DM and DE at the basic
level exists. In this case, addressing the gravitational counterpart may be
an acceptable strategy to repair flaws and address phenomenology. In this
light, expansions and changes to GR appear to be feasible options at both
early and late epochs \cite{I19,I20,I21,I22,II23,II24}. In the context of
GR, the Ricci curvature scalar R generates spacetime dynamics which is
derived from the Levi-Civita connection. Gravity is a metric theory in this
setting, including Lorentz invariance, causality, and other well-established
assumptions. Nevertheless, we may express the gravitational field using
alternative geometric quantities such as torsion and non-metricity \cite%
{I23,I24}. In this direction, an equivalent formulation of GR known as the
symmetric teleparallel equivalent of GR (STEGR) was proposed \cite%
{Ii25,Ii26,Ii27,Ii28}. In this model there the space-time geometry has a
non-metric connection, yet total curvature and torsion are vanishing.
Another possible extension of STEGR is $f(Q)$ gravity \cite{I25}. Such a
theory, which does not need the Equivalence Principle a priori, is
appropriate to be dealt with under the standard of gauge theories and has
additional advantages. For example, it appears to have no major coupling
issues due to extra scalar modes \cite{I26}. Lately, $f(Q)$ gravity has been
seriously considered \cite{I27,I28}, particularly to explain the late-time
acceleration and dark energy difficulties \cite{Ii29,Ii30,Ii31}, although it
has not been thoroughly examined in the early universe.

The purpose of this study is to examine the evolution of the tachyonic
inflation and reheating regimes of the $f(Q)$ gravity model. In the previous
work \cite{IZ}, tachyonic inflation in the $f(Q)$ gravity frame has been
analyzed. However, there was no discussion of the reheating process.
Therefore, we will focus on both tachyonic inflation and reheating on an $%
f(Q)$ gravity model to clarify the physical picture as a way to allow us to
understand the dynamics of inflation along with reheating in the early
Universe. We start by presenting the basic formalism of $f(Q)$ gravity to
construct a tachyonic inflation model based on modifed gravity that could
describe the transition from inflation to reheating later.

After tachyonic inflation, a period occurs in which the stored energy
density in the inflaton is converted to the thermal bath (a plasma of
relativistic particles) \cite{IX,IX1,IXX,IXXX}. This is realized by
reheating, which occurs between the end of inflation and the beginning of
the radiation-dominated epoch. In the process of reheating, ordinary matter
production occurs as a consequence of the energy loss of the\ inflaton
field. In the most basic example, reheating can occur by perturbative decay
of an inflaton into standard model matter particles as the inflaton
oscillates around the minimum of its potential \cite{IX2,IX3}. However, this
scenario has been questioned since it fails to account for the cohesive
character of the inflaton field \cite{IX4,IX5}. In other cases, reheating is
preceded by a preheating phase during which particle creation happens by
nonperturbative mechanisms such as parametric resonance decay \cite{IX6}.
The possibility where preheating and reheating both occur from the end of
inflation to the radiation-dominated epoch was discussed in \cite{I10}. It
is difficult to directly constrain reheating from CMB. Nevertheless,
considering the phase between the time observable scales crossed the horizon
and the present time, we can obtain information via indirect limits. In this
direction, we are concerned to study the reheating phase parametrized by the
number of e-folds $N_{re}$ along with the reheating temperature $T_{re}$ in
a specific interval of the equation of state (EoS) $\omega _{re}.$ We will
constrain the reheating parameters $\left( N_{re},T_{re}\right) $ for the\
tachyonic inflation in $f(Q)$\ gravity according to the recently released
Planck data.

Our paper is organized as follows: In the next section \ref{sec2}, we review
the basic formalism of $f(Q)$\ gravity. In Sect. \ref{sec3}, we develop the
inflationary tachyonic model. In Sect. \ref{sec4}, we present the
inflationary scenario in $f(Q)$\ gravity. In Sect. \ref{sec5}, the
observational constraints are presented. In Sect. \ref{sec6}, we discuss the
reheating senario in tachyonic inflation from an $f(Q)$\ model perspective.
The last section is devoted to a conclusion.

\section{ The basic formalism of $f\left( Q\right) $ gravity}

\label{sec2}

Gravitational effects in Weyl-Cartan geometry are induced not only by a
change in the direction of a vector in parallel transport but also by a
change in its length. Non-metricity describes the variation in the length of
a vector geometrically, and it is mathematically characterized as the
covariant derivative of the metric tensor, which is a generalization of the
gravitational potential.%
\begin{equation}
Q_{\alpha \mu \nu }\equiv \nabla _{\alpha }g_{\mu \nu }=\partial _{\alpha
}g_{\mu \nu }-g_{\nu \sigma }\Sigma {^{\sigma }}_{\mu \alpha }-g_{\sigma \mu
}\Sigma {^{\sigma }}_{\nu \alpha }\neq 0.  \label{2f}
\end{equation}%
In this work, we will consider the modified Einstein-Hilbert action in $f(Q)$
symmetric teleparallel gravity expressed as,%
\begin{equation}
S=\int \frac{1}{2\kappa }f(Q)\sqrt{-g}d^{4}x+S_{\phi },  \label{eqn1}
\end{equation}%
where $f(Q)$ represents a general function of the non-metricity scalar $Q$, $%
g$ is the determinant of the metric tensor $g_{\mu \nu }$ i.e. $g=\det
\left( g_{\mu \nu }\right) $,\ $\kappa =1/M_{p}^{2}$\ and $S_{\phi }$ is the
tachyonic action that we will define in the next section.\ Furthermore, the
non-metricity scalar $Q$ is given as,%
\begin{equation}
Q\equiv -g^{\mu \nu }(L_{\,\,\,\alpha \mu }^{\beta }L_{\,\,\,\nu \beta
}^{\alpha }-L_{\,\,\,\alpha \beta }^{\beta }L_{\,\,\,\mu \nu }^{\alpha }).
\label{eqn2}
\end{equation}%
The non-metricity tensor trace is given by, 
\begin{equation}
Q_{\beta }=g^{\mu \nu }Q_{\beta \mu \nu }\qquad \widetilde{Q}_{\beta
}=g^{\mu \nu }Q_{\mu \beta \nu }.  \label{eqn5}
\end{equation}%
Moreover, the non-metricity scalar $Q$ is connected to the superpotential
tensor (or non-metricity conjugate) $P_{\,\,\,\mu \nu }^{\beta }$, 
\begin{equation}
P_{\,\,\,\mu \nu }^{\beta }=-\frac{1}{2}L_{\,\,\,\mu \nu }^{\beta }+\frac{1}{%
4}(Q^{\beta }-\widetilde{Q}^{\beta })g_{\mu \nu }-\frac{1}{4}\delta _{(\mu
}^{\beta }Q_{\nu )}.  \label{eqn6}
\end{equation}%
The non-metricity scalar in terms of the superpotential tensor is obtained
from the definition above, 
\begin{equation}
Q=-Q_{\beta \mu \nu }P^{\beta \mu \nu }.  \label{eqn7}
\end{equation}

Varying the gravitational action (\ref{eqn1}) with respect to metric tensor $%
g_{\mu \nu }$, we derive the field equations of $f(Q)$ gravity as, 
\begin{equation}
-\frac{2}{\sqrt{-g}}\nabla _{\beta }\left( f_{Q}\sqrt{-g}P_{\,\,\,\,\mu \nu
}^{\beta }\right) -\frac{1}{2}fg_{\mu \nu }-f_{Q}(P_{\mu \beta \alpha
}Q_{\nu }^{\,\,\,\beta \alpha }-2Q_{\,\,\,\mu }^{\beta \alpha }P_{\beta
\alpha \nu })=\kappa T_{\mu \nu }.  \label{eqn11}
\end{equation}%
Here, $f_{Q}=\frac{df\left( Q\right) }{dQ}$, and $\nabla _{\beta }$\
indicates the covariant derivative. Now, we examine a
Friedmann-Lemaitre-Robertson-Walker (FLRW) metric which is homogeneous and
spatially flat. As a result, $Q=6H^{2}$ represents the equivalent
non-metricity scalar. In our current research, we assume that the Universe
is a perfect non-viscosity fluid with the energy-momentum tensor given by 
\begin{equation}
T_{\nu }^{\mu }=diag\left( -\rho ,p,p,p\right) ,  \label{eqn13}
\end{equation}%
where $p$ is the perfect non-viscosity fluid pressure and $\rho $ is the
energy density of the Universe, the modified Friedmann equations are given
as 
\begin{equation}
\kappa \rho =\frac{f}{2}-6FH^{2}  \label{eqn14}
\end{equation}%
and%
\begin{equation}
\kappa p=-\frac{f}{2}+6FH^{2}+2\left( \dot{F}H+F\dot{H}\right) .
\label{eqn15}
\end{equation}%
Here, $\left( \text{\textperiodcentered }\right) $ represents a derivative
with respect to cosmic time $\left( t\right) $, and $F\equiv f_{Q}$
represents differentiation with respect to $Q$. The evolution equation for
the Hubble parameter $H$ can be derived by combining Eqs. (\ref{eqn14}) and (%
\ref{eqn15}) as, 
\begin{equation}
\overset{.}{H}+\frac{\overset{.}{F}}{F}H=\frac{\kappa }{2F}\left( \rho
+p\right) .  \label{eqn16}
\end{equation}%
Einstein's field equations (\ref{eqn14}) and (\ref{eqn15}) can be regarded
as extended symmetric teleparallel equivalents to standard Friedmann's
equations with supplementary terms from the non-metricity of space-time and
the trace of the energy-momentum tensor $T$ which behaves as an effective
component. Therefore, the effective energy density $\rho _{eff}$\ and
effective pressure $p_{eff}$\ are determined by, 
\begin{equation}
3H^{2}=\kappa \rho _{eff}=\frac{f}{4F}-\frac{\kappa }{2F}\rho ,
\label{eqn17}
\end{equation}%
with%
\begin{equation}
2\dot{H}+3H^{2}=-\kappa p_{eff}=\frac{f}{4F}-\frac{2\dot{F}H}{F}+\frac{%
\kappa }{2F}\left( \rho +2p\right) .  \label{eqn18}
\end{equation}%
Taking into account Eqs. (\ref{eqn16}) and (\ref{eqn17}) one gets%
\begin{equation}
\rho =\frac{f-12H^{2}F}{2\kappa }.  \label{eqn19}
\end{equation}%
Before we can check how the inflationary scenario works in $f\left( Q\right) 
$ gravity, we must recall the dynamics of tachyonic inflation.

\section{Tachyonic inflation dynamics}

\label{sec3}

Tachyon inflation is a type of K-inflation model in which inflation is
produced by the use of a noncanonical kinetic term. The tachyon inflation
action is provided by \cite{A1}

\begin{equation}
S_{\phi }=-\int dx^{4}\sqrt{-g}V\left( \phi \right) \left( 1+g^{\mu \nu
}\partial _{\mu }\phi \partial _{\nu }\phi \right) ^{\frac{1}{2}},
\end{equation}%
in a spatially flat FRW Universe, a rolling tachyon condensate can be
characterized by an effective fluid with the energy-momentum tensor \cite%
{A2,A3}, where the energy density and pressure for the background part of
the tachyon field are given as%
\begin{eqnarray}
\rho &=&\frac{V\left( \phi \right) }{\left( 1-\dot{\phi}^{2}\right) ^{\frac{1%
}{2}}},  \label{eqn20} \\
p &=&-V(\phi )\left( 1-\dot{\phi}^{2}\right) ^{\frac{1}{2}},  \label{eqn21}
\end{eqnarray}%
here $\phi $ is the tachyon field with length dimension and $V\left( \phi
\right) $ represents its potential. Several potential choices have been
obtained using string theory \cite{A4,A5,A6,A7,A8}. In this work we consider
the exponential potential \cite{A9,A10} given by%
\begin{equation}
V\left( \phi \right) =\lambda \exp \left( -\frac{\phi }{\phi _{0}}\right) .
\label{exp}
\end{equation}%
Hence the Friedmann equation takes the form%
\begin{equation}
H^{2}=\frac{\kappa }{3}\rho \equiv \frac{\kappa }{3}\frac{V\left( \phi
\right) }{\left( 1-\dot{\phi}^{2}\right) ^{\frac{1}{2}}}.
\end{equation}%
The Friedmann equation could be used to calculate the conditions to achieve
inflation which is given as 
\begin{equation}
\frac{\ddot{a}}{a}=-\frac{\kappa }{6}\left( \rho +3p\right) =\frac{\kappa }{3%
}\frac{V\left( \phi \right) }{\left( 1-\dot{\phi}^{2}\right) ^{\frac{1}{2}}}%
\left( 1-\frac{3}{2}\dot{\phi}^{2}\right) >0,
\end{equation}%
that gives $\dot{\phi}^{2}<2/3.$\ The tachyon field equation of motion
considering a minimall coupling to gravity is obtained as%
\begin{equation}
\frac{\ddot{\phi}}{\left( 1-\dot{\phi}^{2}\right) }+3H\dot{\phi}+\ln \left(
V\right) ^{\prime }=0.
\end{equation}%
For inflation to persist long enough, $\ddot{\phi}$ should be less than the
friction term in the tachyon field equation of motion $\ddot{\phi}<3H\dot{%
\phi}$\ 
\begin{eqnarray}
\dot{\phi} &\sim &-\frac{1}{3H}\frac{V^{\prime }\left( \phi \right) }{%
V\left( \phi \right) }, \\
H^{2} &\sim &\frac{V(\phi )}{3M_{p}^{2}}.
\end{eqnarray}%
The slow-roll parameters can be constructed in terms of the Hubble parameter
as \cite{A11} 
\begin{eqnarray}
\epsilon &=&-\frac{\dot{H}}{H^{2}},  \label{eps1} \\
\eta &=&\frac{\dot{\epsilon}}{H\epsilon }\approx -\frac{\ddot{H}}{2\dot{H}H}.
\label{eps2}
\end{eqnarray}%
The number of e-folds associated with inflation is defined as follows 
\begin{equation}
N\equiv \ln \left( \frac{a_{end}}{a}\right) =\int_{t}^{t_{end}}Hdt,
\end{equation}%
the index $"_{end}"$ denotes the time when inflation ended. Inflation will
continue as long as $\epsilon _{i}<1$\ to solve the standard cosmological
problems. However, at the end of inflation, the slow-roll parameter must
reach $\epsilon _{i}=1.$ The relation between the curvature perturbations
and the slow roll parameters is given by the spectral index $n_{s}$. An
additional parameter used to study the period of inflation is the e-folding
number which describes the rate of the Univers expansion during this period,
the tensor to scalar perturbations ratio, $r,$ and the spectral index\ $%
n_{s} $ are given in the following way \cite{A1}%
\begin{eqnarray}
n_{s}-1 &=&-2\epsilon -2\eta ,  \label{ns} \\
r &=&16\epsilon .  \label{r}
\end{eqnarray}%
Next, we will discuss how the tachyonic inflationary scenario can be
described in the context of $f(Q)$ gravity.\ \ 

\section{ Tachyonic inflationary scenario in $f(Q)$ gravity}

\label{sec4}

In the context of cosmic inflation, we consider the model $f(Q)=\alpha Q,$
with $\alpha =F\neq 0$, by taking Eqs.\ (\ref{eqn16}), (\ref{eqn17}), (\ref%
{eqn18}) and (\ref{eqn19})\ we obtain the following equations%
\begin{eqnarray}
\dot{H} &=&\frac{\kappa \rho \left( 1+\omega \right) }{2\alpha },  \label{Hp}
\\
3H^{2} &=&\kappa \rho _{eff}=-\frac{\kappa \rho }{\alpha },  \label{eqnH} \\
2\dot{H}+3H^{2} &=&-\kappa p_{eff}=\frac{\kappa \omega \rho }{\alpha },
\label{eqnHH} \\
\rho &=&-\frac{3\alpha H^{2}}{\kappa }.
\end{eqnarray}%
\newline
In of $f(Q)$ gravity, we consider the tachyonic field leads from Eqs. (\ref%
{eqn20}), (\ref{eqn21}) along with Eqs.(\ref{eqn17}), (\ref{eqn18}) gives%
\begin{eqnarray}
3H^{2} &=&-\frac{\kappa }{\alpha }\frac{V}{\left( 1-\dot{\phi}^{2}\right) ^{%
\frac{1}{2}}},  \label{pp} \\
2\dot{H}+3H^{2} &=&-\frac{\kappa }{\alpha }V\left( 1-\dot{\phi}^{2}\right) ^{%
\frac{1}{2}}.  \label{p}
\end{eqnarray}%
Taking into account the previous two equations one gets%
\begin{equation}
\dot{H}=\frac{\kappa V}{2\alpha }\frac{\dot{\phi}^{2}}{\left( 1-\dot{\phi}%
^{2}\right) ^{\frac{1}{2}}},  \label{H.}
\end{equation}%
after the derivation of Eq.(\ref{eqnH}) and using Eq. (\ref{eqnHH}), we
obtain the modified equation of motion%
\begin{equation}
\frac{\ddot{\phi}}{\left( 1-\dot{\phi}^{2}\right) }+3H\dot{\phi}+\frac{%
V^{\prime }}{V}=0.
\end{equation}%
From Eqs. (\ref{eps2}) we know that the condition $\eta \ll 1$ leads to $%
\ddot{\phi}\ll 3H\dot{\phi},$\ which means that the time derivative of the
tachyonic field is given as%
\begin{equation}
\dot{\phi}=-\frac{1}{3H}\frac{V^{\prime }}{V}.  \label{phip}
\end{equation}

Now, let us use the slow-roll parameters that will be calculated in the
context of the $f(Q)$ gravity. For this purpose, we should insert Eqs. (\ref%
{pp}) and (\ref{H.}) into the slow roll parameter, knowing that we need to
consider the case $\dot{\phi}^{2}<2/3$ which gives%
\begin{eqnarray}
\epsilon &\approx &\frac{3}{2}\dot{\phi}^{2}\simeq -\frac{\alpha }{2\kappa }%
\left( \frac{V^{\prime }}{V}\right) ^{2}\frac{1}{V}, \\
\eta &\approx &-\frac{\ddot{\phi}}{H\dot{\phi}}\simeq -\frac{\alpha }{\kappa 
}\left( \frac{V^{\prime \prime }}{V}\right) \frac{1}{V},
\end{eqnarray}%
the e-folds number for this chosen model can be written using Eq.(\ref{pp})
and Eq.(\ref{phip}) in the following way{}%
\begin{equation}
N\equiv \int_{\phi _{k}}^{\phi _{end}}\frac{H}{\dot{\phi}}d\phi \approx
-\int_{\phi _{end}}^{\phi _{k}}\frac{\kappa }{\alpha }\frac{V^{2}}{V^{\prime
}}d\phi .  \label{N}
\end{equation}%
Our goal for the next section is to calculate the inflationary tachyonic
parameters for the exponential model.

\section{Observational constraints}

\label{sec5}

\subsection{Exponential model}

The exponential potential for tachyon inflation plays an important role
within the inflationary cosmology \cite{E1,E2}. Taking into account the
parametrization $\varphi =\phi /\phi _{0}$ Eq.(\ref{exp}) is given in the
form%
\begin{equation}
V\left( \varphi \right) =\lambda \exp \left( -\varphi \right) ,
\end{equation}%
for simplicity, we define a dimensionless constant as $\Phi _{0}=\lambda
\kappa \phi _{0}^{2}/\alpha ,$\ the slow roll parameters $\epsilon $ and $%
\eta $\ for this model are expressed as\ 
\begin{eqnarray}
\epsilon &\approx &-\frac{1}{2\Phi _{0}}\exp \left( \varphi \right) ,
\label{Slow1} \\
\eta &\approx &-\frac{1}{\Phi _{0}}\exp \left( \varphi \right) .
\label{Slow2}
\end{eqnarray}%
The value of the field at the end of inflation is obtained considering $%
\epsilon _{end}=1$%
\begin{equation}
\exp \left( \varphi _{end}\right) =-2\Phi _{0},
\end{equation}%
from the inflationary number Eq.(\ref{N}) we calculate the following
expression%
\begin{equation}
N\approx -\Phi _{0}\left( \exp \left( -\varphi _{k}\right) -\exp \left(
-\varphi _{end}\right) \right) =-\Phi _{0}\left( \exp \left( -\varphi
_{k}\right) +2\Phi _{0}\right) ,
\end{equation}%
finally, when the mode $k$ leaves the horizon, the tachyon field value $%
\varphi _{k}$\ is given as%
\begin{equation}
\exp \left( -\varphi _{k}\right) =-\frac{N+\frac{1}{2}}{\Phi _{0}}.
\label{Phik}
\end{equation}%
\ Here, we must mention that for a mathematically consistent equation it is
necessary to consider the bounds $\Phi _{0}<0.$\ Moreover, using the slow
roll parameters Eqs.(\ref{Slow1}) and (\ref{Slow2}), along with Eq. (\ref{ns}%
), the observable quantities can be obtained in terms of $N$ as%
\begin{eqnarray}
1-n_{s} &=&\frac{2}{N+\frac{1}{2}},  \label{nsN} \\
r &=&\frac{8}{N+\frac{1}{2}}.
\end{eqnarray}%
From the previous two equations, we conclude that the choice of tachyonic
exponential inflation allows us to find an analytic relation between $r$, $%
n_{s}$ independently of $\Phi _{0}$\ parameter. The $N$ can be expressed in
terms of $n_{s}$ for the chosen exponential model in $f(Q)$ gravity, which
from Eq. (\ref{nsN}) we can obtain%
\begin{equation*}
N=\frac{2}{1-n_{s}}-\frac{1}{2}.
\end{equation*}%
\begin{figure}[tbp]
\centering
\includegraphics[width=16cm]{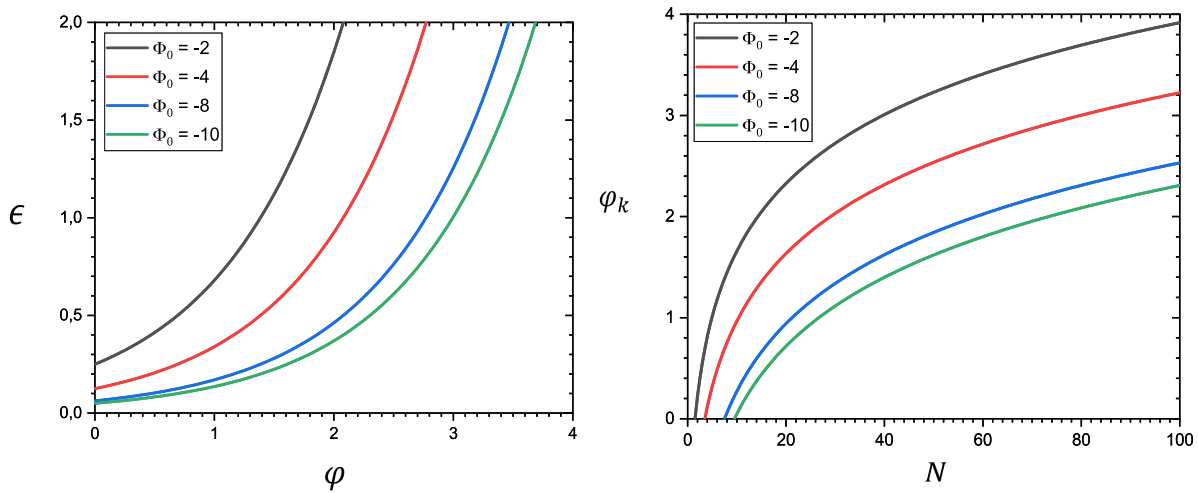}
\caption{In the left panel, we study the behavior of $\protect\epsilon $ as
a function of $\protect\varphi $ for different values of $\Phi _{0},$ while
in the right one, we plotted the evolution of the field at the beginning of
inflation $\protect\varphi _{k}$\ as functions of the inflationary e-folds.
For both plots, we choose the exponential model for negative values of $\Phi
_{0}.$}
\label{fig:1}
\end{figure}
In Fig. (\ref{fig:1}) we provide a numerical evaluation of the slow-roll
parameter $\epsilon $ as a function of the field parameter $\varphi ,$ along
with the evolution of the field parameter $\varphi _{k}$\ at the beginning
of inflation as functions of the e-folds number $N,$\ and proved that for a
mathematical consistency $\Phi _{0}$ must be negative, $\Phi _{0}=-2$\
represents the black line, $\Phi _{0}=-4$\ represents the red line, $\Phi
_{0}=-8$\ represent the blue line and $\Phi _{0}=-10$ is presented with the
green one. At the end of inflation when $\epsilon _{end}=1$\ the field
parameter values increases as we decrease $\Phi _{0}$\ from $-2$\ to $-10.$\
While the field parameter increases rapidly as functions of the e-folds
number take higher values of $\Phi _{0}.$%
\begin{figure}[h]
\centering
\includegraphics[width=16cm]{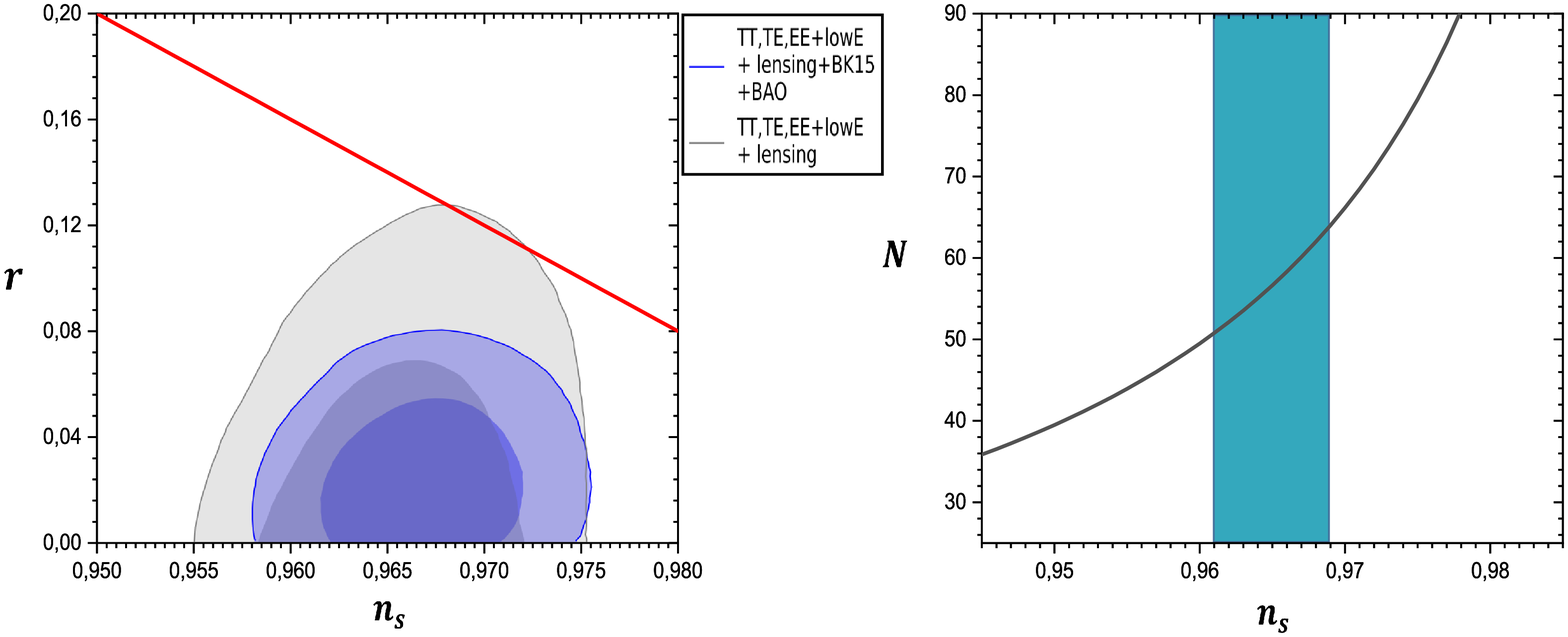}
\caption{On the left side $r$ as a function of $n_{s}$ is plotted for
exponential tachyonic inflation in $f(Q)$ gravity. Inner and outer shaded
regions are $1\protect\sigma $ and $2\protect\sigma $ constraints in
combination with CMB lensing reconstruction and BAO from Planck data,
respectively. On the right side, the inflationary e-folds evolution as
fictions of $n_{s}$\ is presented.}
\label{fig:2}
\end{figure}
In Fig. (\ref{fig:2}) we plot the behavior of the tensor-to-scalar-ratio as
a function of the spectral index $n_{s},$ the decreasing function provides
consistency with observations for a specific interval according to recent
observations. As for the evolution of the inflationary e-folds, $N$ must be
bounded as $50\leq N\leq 64$\ to reproduce the observational bound on $n_{s}$%
.

Now, as we constrain our inflationary parameters, we are ready to study the
reheating phase for tachyonic inflation in $f(Q)$ gravity.

\section{Reheating after tachyonic inflation in $f(Q)$ gravity}

\label{sec6}

When the equation-of-state parameter reaches the value $\omega =1/3$,
inflation terminates. In~standard inflationary cosmology we~assume~that,
following inflation, the universe goes through a reheating phase in which
the inflaton field coherently oscillates at the minimum of its potential,
converting its energy to the relativistic created particles. Reheating
models can be parameterized in terms of the thermalization temperature $%
T_{re}$ at the end of the reheating, the duration of the reheating $N_{re}$,
and a constant equation of state during the reheating era $\omega _{re}$ 
\cite{F1,F2}. Following the approach presented in \cite{F3,F4,F5,F6,F7,F8}
where the current observations are considered to be linked to the evolution
of the inflaton field during inflation taking into account the relation
between the horizon exit of the comoving Hubble radius $a_{k}H_{k}=k$, and
the present time

The reheating duration can be extracted considering the phase between the
time observable CMB scales crossed the horizon and the present time.
Deferent eras occurred throughout this length of time that can be described
by the two following equations \cite{F9}:

\begin{equation}
\frac{k}{a_{0}H_{0}}=\frac{a_{k}}{a_{end}}\frac{a_{end}}{a_{re}}\frac{a_{re}%
}{a_{eq}}\frac{a_{eq}H_{eq}}{a_{0}H_{0}}\frac{H_{k}}{H_{eq}},  \label{Eq1}
\end{equation}%
the pivot scale for a specific experiment is parametrized by $k$\ \cite{F9},$%
\ N$ is the $e$-folds of the inflation era, $N_{re}$ and $N_{RD}$
respectively correspond to the reheating and radiation-domination era
durations. Using $\rho \propto a^{-3(1+\omega ^{\ast })}$, the reheating
epoch is described by :%
\begin{equation}
\frac{\rho _{end}}{\rho _{re}}=\left( \frac{a_{end}}{a_{re}}\right)
^{-3(1+\omega _{re})},  \label{Eq2}
\end{equation}%
where $\rho _{end}$ and $a_{end}$ correspond to the end of inflation and $%
\rho _{re}$ and $a_{re}$ correspond to the end of reheating. As a result :%
\begin{equation}
N_{re}=\frac{1}{3\left( 1+\omega _{re}\right) }\ln \left( \frac{\rho _{end}}{%
\rho _{re}}\right) ,  \label{Eq3}
\end{equation}%
knowing that the energy density at reheating is given by%
\begin{equation}
\rho _{re}=\frac{\pi ^{2}g_{re}}{30}T_{re}^{4},  \label{Eq4}
\end{equation}%
with%
\begin{equation}
T_{re}=\left( \frac{43}{11g_{re}}\right) ^{\frac{1}{3}}\frac{a_{0}}{a_{re}}%
T_{0}.  \label{Eq5}
\end{equation}%
The reheating e-folds number is obtained from Eq. (\ref{Eq3}) and Eqs. (\ref%
{Eq4}) with (\ref{Eq5}) as%
\begin{equation}
N_{re}=\frac{4}{1-3\omega _{re}}\left[ 6.61-\ln \left( \frac{V_{end}^{\frac{1%
}{4}}}{H_{k}}\right) -N\right] .  \label{Eq6}
\end{equation}%
the previous expression was obtained considering $g_{re}\approx 100$\ and
the pivot scale $0.05Mpc$, and $M_{pl}=2.435\times 10^{18}GeV$, $a_{0}=1$, $%
T_{0}=2.725K$ values from \cite{F9}. The reheating temperature $T_{re}$%
\bigskip\ is then given by%
\begin{equation}
T_{re}=\left[ \left( \frac{43}{11g_{re}}\right) ^{\frac{1}{3}}\frac{%
a_{0}T_{0}}{k}H_{k}~e^{-N_{k}}\left( \frac{36V_{end}}{\pi ^{2}g_{re}}\right)
^{\frac{-1}{3(1+\omega _{re})}}\right] ^{\frac{3(1+\omega _{re})}{3\omega
_{re}-1}}.  \label{Eq7}
\end{equation}

This section's primary conclusions are the expressions for the reheating
temperature $T_{re}$ Eq. (\ref{Eq7}) and the reheating of e-folds number Eq.
(\ref{Eq6}) which rely on three main parameters $H_{k}$, $N$, and $V_{end}$.
Next, we obtain these parameters for tachyon inflation with exponential
potential in terms of the spectral index ns and the amplitude of scalar
perturbations $A_{s}$. With this, we constrain the reheating temperature and
duration for the tachyon inflation in $f(Q)$ gravity considering $\omega
_{re}$ to be between $-1/3$ and $1/4$.

At the end of inflation, the potential for the exponential model is
expressed using the expression of $V_{k}$\ as%
\begin{equation}
V_{k}=-\frac{3\alpha }{\kappa }H_{k}^{2}=\lambda \exp \left( -\varphi
_{k}\right) ,  \label{Eq8}
\end{equation}%
which gives 
\begin{equation}
V_{end}=\frac{3H_{k}^{2}}{\kappa }\frac{\exp \left( -\varphi _{end}\right) }{%
\exp \left( -\varphi _{k}\right) }=-\frac{3\alpha }{\kappa }H_{k}^{2}\left(
1-n_{s}\right) .
\end{equation}%
Following the results of the previous section, here we must mention that as
the parameter $\Phi _{0}=\lambda \kappa \phi _{0}^{2}/\alpha $ must be $\Phi
_{0}<0,$\ this directly leads to consider the bound $\alpha <0.$ Next, at
the pivot scale, we obtain $P_{\zeta }=A_{s}$ and $r=P_{h}/P_{\zeta }$ (with 
$P_{h}=2\kappa H_{k}^{2}/\pi ^{2}$), the expression of Hubble constant at
the time of horizon exit of mode $k$ which can be written in terms of scalar
amplitude $A_{s}$ and spectral index $n_{s}.$ 
\begin{equation}
H_{k}=\sqrt{2\pi ^{2}\frac{A_{s}}{\kappa }\left( 1-n_{s}\right) }.
\end{equation}%
\begin{figure}[tbp]
\centering
\includegraphics[width=16cm]{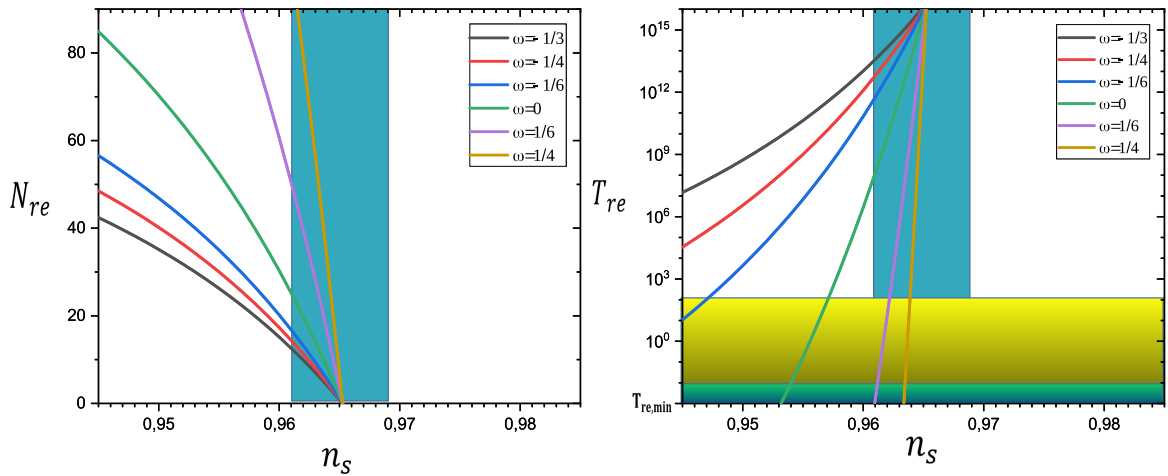}
\caption{Constraints on the reheating duration $N_{re}$ \ and temperature $%
T_{re}$\ for the tachyonic potential in $f(Q)$\ gravity with different
values of the reheating equation of state which we choose to be bounded as $%
-1/3\leq \protect\omega _{re}\leq 1/4.$}
\label{fig:3}
\end{figure}

In Fig. (\ref{fig:3}), constraints on the reheating duration $N_{re}$ and
temperature $T_{re}$\ for the tachyonic potential in $f(Q)$ gravity with
different values of the reheating equation of state bounded as $-1/3\leq
\omega _{re}\leq 1/4$ are presented, in the case of the tachyonic inflation
in $f(Q)$ gravity. The vertical blue region represents Planck's bounds on $%
n_{s}=0.9649\pm 0.0042$ \cite{F10}. Reheating occurs instantaneously at the
point where all the lines converge, we define it by the limit $%
N_{re}\rightarrow 0$. We observe that this model does not depend on the
values of $\alpha $ from Eq. (\ref{Eq8}). Additionally to that, all the
lines are shifted towards the spectral index central value, which means
reheating duration show good consistency for a wide range of $N_{re}$ value.
On the other hand, the reheating temperature $T_{re}$ converges towards $%
10^{16}GeV$ where the instantaneous reheating is defined. Temperatures below
the green region are ruled out by BBN, and the yellow region represents $%
100GeV$ of the electroweak scale. This model predict also a wide range of
reheating temperatures and durations as we increase the EoS towards $\omega
_{re}\rightarrow 1/4$ according to recent Plank's observations.

\section{Conclusion}

\label{sec7}

In this paper, we have reviewed tachyonic inflation and reheating in $f(Q)$
gravity in which we constrain the inflationary parameters and the reheating
parameters according to recent observations.\ We first present the formalism
of tachyon inflation in our chosen modified $f(Q)$ model where we found
that\ the slow roll and e-folds number corresponding to inflation depends on
the $f(Q)$ model $\alpha $\ parameter. We have found that, an exponential
potential of tachyonic inflation present consistency for specific ranges in
the $(r,ns)$ plane independently of $f(Q)$ model parameters. While the
inflationary e-folds, $N$ must be bounded as $N\in \left[ 50,64\right] $\ to
reproduce the observational bound on $n_{s}.$ Moreover, we have also
calculated the reheating duration $N_{re}$\ and temperature $T_{re}$\ \ for
the exponential potential in $f(Q)$ gravity as a function of spectral index $%
n_{s}$ by considering a constant equation of state during reheating $\omega
_{re}$ and considering some specific values in the interval $\omega _{re}\in %
\left[ -1/3,1/4\right] $. For the modified $f(Q)$ model, the exponential
potential of tachyonic inflation predicts consistent values of reheating
duration and temperature knowing that the ranges of $N_{re}$ and $T_{re}$\
get larger when we increase the values of the EoS towards $\omega
_{re}\rightarrow 1/4$ according to recent Planck's observations.

\end{document}